**AC losses in high pressure synthesized $MgB_2$ bulk rings measured by transformer method**

V. Meerovich[a], V. Sokolovsky[a], T. Prikhna[b] , W. Gawalek[c], and T. Habisreuther[c]

[a] *Physics Department, Ben-Gurion University of the Negev, Beer-Sheva, 84105, Israel*
[b] *Institute for Superhard Materials of the National Academy Sciences of Ukraine, 2, Avtozavodskaya St., Kiev, 04074, Ukraine*
[c] *Institut fur Photonicsche Technologien, Albert-Einstein Strasse 9, Jenna, D-07745, Germany*

**Abstract:**

The recently developed manufacturing technologies use high pressure and various doping additions to prepare bulk $MgB_2$-based materials with a high critical current density measured by the magnetization methods. We use a contactless transformer method, which is based on studying the superconductor response to an induced transport current, to measure AC losses in bulk $MgB_2$ rings synthesized under high pressure. The obtained dependence of the losses on the primary current (applied magnetic field) is fitted by the power law with the exponent ~2.1 instead of the cubic dependence predicted by Bean's model and power-law electric field-current density (E-J) characteristics with a large exponent. The unusually strong dependence of AC losses on the frequency was also observed. It was shown that the E-J characteristic of bulk $MgB_2$ is well fitted by the dependence used in the extended critical state model based on account of the viscous vortex motion in the flux flow regime. Numerical simulation using this E-J characteristic gives the current and frequency AC loss dependences which well agree with the experimental results.

**Keywords:** superconductor bulk $MgB_2$, AC losses, voltage-current characteristic

## 1. Introduction

It is expected that bulk $MgB_2$-based materials can successfully compete with high-temperature bulk superconductors in fault current limiters, motors, generators, superconducting magnets, and etc. [1-7]. The recently developed manufacturing technologies use high pressure and various doping additions to prepare bulk $MgB_2$ materials with a high critical current density $J_c$. The evaluations of $J_c$ using the results of the magnetization measurements and the critical state model give the values of $10^6$ A/cm$^2$ at 20 K in the magnetic field up to 1 T [3, 8-10].

The problem of AC loss evaluation is an important issue for the development of the superconducting devices for applications under AC conditions. AC losses determine ranges of the rated currents and magnetic fields for the devices, required power of cryogenic equipment and economical gain. Other aspects of AC losses are connected with studying the physical properties of superconductors so as: microscopic motion of the Abrikosov vortices, phase state of the vortex lattice and etc.

In this paper we report the results of measuring AC losses in $MgB_2$ rings synthesized under high pressure. In contrast to the magnetization methods, we apply a contactless transformer method which allows us to study the characteristics of a superconductor with an induced transport current.

Our experimental method is described in detail in the next section. Description of the experimental set-up and tested samples is given in section 3. In section 4 we report the experimental results. The obtained dependence of the losses on the primary current is fitted by a power law with the exponent ~2.1 instead of the cubic dependence predicted by Bean's model. The unusually strong dependence of the AC losses on the frequency is also presented. In section 5 we discuss the obtained results and show that the observed AC loss behavior is expected for samples with the E-J characteristic (E – electric field, J – current density) described by the extended critical state model based on taking into account of the viscous vortex motion in the flux flow regime.

## 2. Experimental method

The measurements of AC losses in bulk superconducting samples with a transport current are usually based on the four-point method which requires a high current supply up to tens or even hundreds of thousands of amperes and high quality contacts. One contactless method suitable for samples in the form of a closed loop (hollow cylinder, ring, or short-circuited coil) is based on using the transformer configuration (Fig. 1) [11, 12]. A superconducting closed loop forms the secondary coil of a transformer in which the primary coil is connected with an AC source. To increase the coupling between the coils, they are centered on a ferromagnetic core in the form of a laminated rod (open core design). The procedure of the AC loss determination in samples with induced current relies on measuring, with the help of the Hall-probe technique (Fig. 1), the magnetic flux density as a function of the instantaneous current in the primary coil. This method has been previously used by the authors of the present work to measure AC losses in BSCCO cylinders [11, 12]. The main advantages of this method are: i) the high currents up to several tens of thousands amperes can be achieved in a superconductor using usual laboratory equipment; ii) no current terminals and measuring contacts to a superconductor are required; iii) measurement results are not influenced by heating due to losses in contacts.

To clarify the basics of the method, let us use the well-known equations of an ideal two-coil transformer

$$u_1 = ri_1 + L_1 \frac{di_1}{dt} + M \frac{di_{sc}}{dt} \quad (1)$$

$$0 = u_{sc} + L_2 \frac{di_{sc}}{dt} + M \frac{di_1}{dt} \quad (2)$$

where $u_1$ is the voltage drop across the primary coil, $u_{sc}$ is the voltage appearing across the superconductor, secondary short closed coil, $i_1$ and $i_{sc}$ are the currents in the primary and secondary coils, and $L_1$ and $L_2$ are their inductances, respectively, $M$ is the mutual inductance of the coils.

The Hall probe located on the top of the rod gives the total magnetic field $B_\Sigma = B_1+B_2$, where $B_1$ and $B_2$ are the magnetic fields produced by the primary and secondary currents, respectively. These fields have opposite directions and proportional to the currents in the coils: $B_1 = k_{b1}i_1N$ and $B_2 = k_{b2}i_{sc}$ ($N$ is the turn number in the primary coil).

Therefore, $B_\Sigma = k_{b1}i_1N + k_{b2}i_{sc}$ and the current in the superconductor is

$$i_{sc} = \frac{1}{k_{b2}}\left(B_\Sigma - k_{b1}Ni_1\right). \quad (3)$$

The voltage drop in the ring is determined from (2) with substitution of (3):

$$u_{sc} = \left(\frac{k_{b1}NL_2}{k_{b2}} - M\right)\frac{di_1}{dt} - \frac{L_2}{k_{b2}}\frac{dB_\Sigma}{dt}. \quad (4)$$

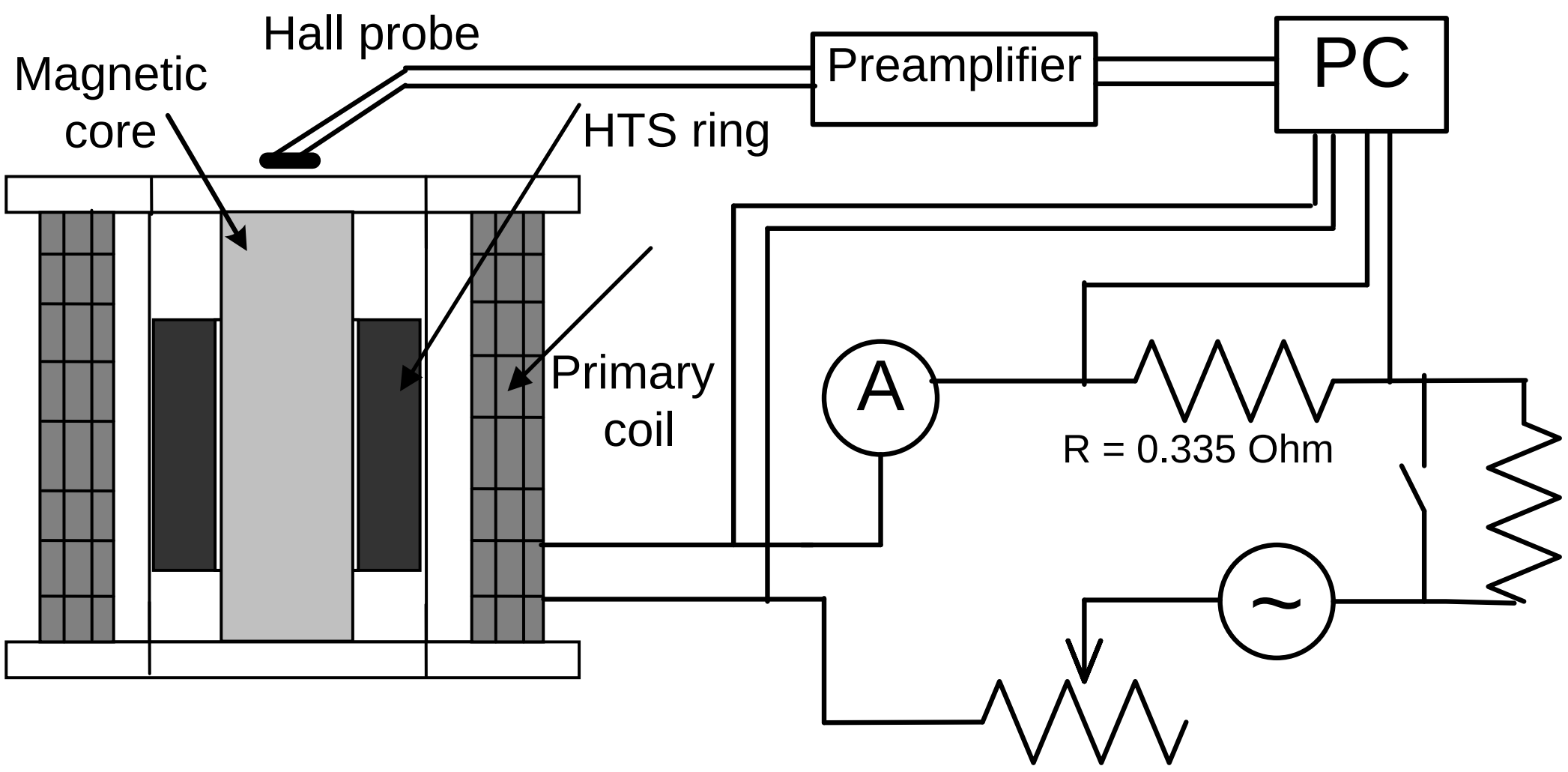

Fig. 1. Set-up for measurements of AC losses by a contactless transformer method.

The expression for AC loss power can be obtained by multiplying (3) by (4) and integrating over the current period $T$. If $k_{b1}$ and $k_{b2}$ are independent of the currents, we have:

$$P = \frac{1}{T}\int_0^T u_{sc} i_{sc} \mathrm{d}t = -\frac{M}{k_{b2} T} \oint B_\Sigma \mathrm{d} i_1 \,. \qquad (5)$$

The last expression shows that AC loss values are proportional to the area of the hysteresis loop on the $B_\Sigma - i_1$ plane.

The right-hand side of (5) contains only directly measured quantities $B_\Sigma$ and $i_1$ . In contrast to the difference methods, the measurement accuracy does not depend on the value of the losses in the primary coil. Errors in determination of $M$ and $k_{b2}$ affect only the magnitude of losses but not the dependence of losses on the current (magnetic field) and frequency.

## 3. Experimental samples and set-up

### *3.1 Experimental samples*

The samples were synthesized under quasihydrostatic high pressure conditions: 2 GPa, 1050 $^{o}$C during 1 hour, from the mixture of Mg and B powders with 4.0 μm average grain size taken in $MgB_2$ stoichiometry with additions of 10 % SiC.

The initial amorphous boron contained 0.94 % C, 0.35% N and 0.72% H (ring 1) or 0.34% N, 0.47% C, 0.37% H, 1.5% O (ring 2). Boron was mixed and milled with Mg chips in high speed planetary activator and then 200-800 nm SiC granules have been added. The technology was described in details in [13]. The rings with outer diameter of 24.3 mm, height of 7 mm (ring 1) and 7.7 mm (ring 2) and wall thickness of 3.2 mm were cut out from high-pressure synthesized blocks. From the magnetization experiments, the critical temperature of the $MgB_2$ samples was estimated to be 39 K.

*3.2 Experimental set-up*

The experimental set-up included a two-coil transformer where the $MgB_2$ ring forming a single-turn secondary coil was placed inside a cooper 700 turn primary coil (Fig. 1). The primary normal metal coil with the internal diameter of 25 mm and the height of 32 mm was wound from 0.18 mm diameter cooper wire. The ferromagnetic laminated open core assembled from 0.3 mm transformer steel sheets had the effective diameter of 15 mm and the height of 48 mm and was inserted into the ring.

The whole device was placed into a cryostat with liquid helium at 4.2 K. The primary coil was connected in series to a circuit including a generator of AC voltage, two variable resistors, and a resistive shunt of 0.335 Ohm for the current measurement. A Hall probe with the sensitivity of 100 mV/T at 4.2 K was placed at the center of the top of the ferromagnetic rod (Fig. 1). The voltage generated by the Hall probe was amplified by a factor 100 and acquired along with the trace of the primary current and voltage drop across the primary coil by a multichannel data acquisition device. The recording rate was 500,000 samples/sec per channel. To prevent heating of a ring due to AC losses, a special electronic switch short-circuited one of the resistors for several AC periods thus providing high currents only for ~ 0.1 s. The pause between measurements was about 15 min to provide cooling the ring till the bath temperature.

*3.3. Determination of the transformer device parameters*

In order to determine the parameters included in Eqs. (1)-(5), the preliminary measurements were performed. From DC experiment, the resistance of the primary coil at 300 K (room temperature) was found to be 66 Ohm and it drops to 0.35 Ohm at 4.2 K.

The main parameters of a transformer are experimentally determined from the open and short circuit tests. The inductance $L_1$ = 0.025 H is obtained from the AC measurement without a ring (the open circuit test). The same inductance was obtained with a ring at 77 K that indicates that the superconductor resistance in the normal state is much larger than the transformer impedance.

The proportionality coefficient $k_{b1}$ was also obtained in the open circuit test, without a secondary coil, when the primary coil is connected with DC source. Our experiments demonstrated that the inductance and $k_{b1}$ do not depend on the magnitude of the primary current, in complete agreement with the well-known linearity of open-core magnetic systems [14, 15].

To determine $k_{b2}$ , we calculated the magnetic field distribution in device with help of COMSOL (COMSOL Multiphysics is a commercial FEM software) using the 2D axially symmetric model in which the rod was replaced by a ferromagnetic lossless cylinder with relative magnetic permeability ranging from 100 to 10.000. The superconducting ring was simulated by a single-turn coil of the same height and external diameter. To simulate various penetration depth of magnetic field into a superconductor the coil thickness was varied from 0.5 mm till 3.2 mm. The simulation with the use of COMSOL program shows that $k_{b1} = k_{b2}$ within the accuracy of 1%. This is explained by the choice of the point for the field measurement at the center of the rod top where the field depends only on the ampere-turns and not on the dimensions of the coil producing the field. Therefore, $B_{\Sigma} = k_b\,(i_1N + i_{sc})$, where $k_b$ is the general notation for $k_{b1}$ and $k_{b2}$.

The mutual inductance $M$ can be found from the short circuit test at low primary currents when the AC losses can be neglected. Actually, the voltage drop across a type-II superconductor under AC conditions is determined by the electric field $E$ induced by the magnetic field $H$ penetrating into the superconductor. From the Faraday equation $\mathrm{rot}\vec{E} = -\mu_0 \frac{\partial \vec{H}}{\partial t}$ in the 1D case, the maximum electric field is estimated as $E \approx \omega\mu_0 H_0 \Delta$, where $H_0$ is the magnetic field intensity amplitude, $\mu_0$ is the vacuum magnetic permeability; $\omega = 2\pi f$, $f$ is the magnetic field frequency. The magnetic field penetration depth $\Delta$ can be estimated using the Bean model $\Delta = H_0 / J_c$ ($J_c$ is the critical current density). Therefore, the maximum of the electric field and voltage drop $u_s$ decrease proportional to the square of the amplitude of the applied magnetic field or current, and at a low primary current, $u_s$ in Eq. (2) can be neglected. Under short circuit tests the voltage drop across the primary coil is determined, on the one hand, by Eq. (1) and, on the other hand, can be expressed as

$$u = ri_1 + L_s \frac{\partial i_1}{\partial t} \,. \qquad (6)$$

From Eqs. (1), (2) and (6) at $u_s$=0 :

$$M = \sqrt{(L_1 - L_s)L_2} \qquad (7)$$

where $L_s$ = 0.0067 H is the inductance of the primary coil measured at low currents (less than 0.2 A) and at the temperature of 4.2 K when the losses in the superconducting ring are negligible (the short circuit test).

At a low primary current a superconducting current is proportional to the primary current. The Hall probe registers a finite magnetic field proportional to the primary current. This effect is explained by finite sizes of the primary coil, ring and magnetic core, and, hence, imperfect magnetic coupling between the coil and the ring. From Eqs. (2) and (3) we obtain

$$L_2 = \frac{M}{N\left(1 - k_b^{'} / k_{b1}\right)} \qquad (8)$$

where $k_b'$ is the proportionality coefficient between $B_\Sigma$ and $Ni_1$ measured at a low primary current. Eqs. (7) and (8) determine the mutual inductance $M$.

*3. 4. Influence of a ferromagnetic core*

All expressions above were obtained assuming that the transformer is ideal. In reality, a transformer with a ferromagnetic core has hysteresis and eddy-current losses. The losses determined by expression (5) include also the losses in the core. We use a laminated core in the form of a rod made of thin, insulated soft iron sheets (open core design) providing high linearity of the B-H characteristic of the magnetic system and low hysteresis losses.

Let us consider the contribution of eddy current losses. As shown in [15], the losses in a thin sheet are mostly due to the normal component of the magnetic field (i.e. the radial component of magnetic field in our device). The radial component at the lateral surface of the ferromagnetic rod was calculated using COMSOL. A ring in the superconducting state was simulated by single-turn coil with resistivity of $10^{-12}$ Ω·m. Fig. 2 shows the distribution of the magnetic flux density $B_r$ at the lateral surface of the core vs. the $z$-coordinate directed along the device axis, calculated for the current of 1 A in a 700-turn primary coil. The maximum of the radial field $B_r$ in the device without a ring is about 0.035 T while, in the device with the ring, it does not exceed 0.003 T. The obtained field distribution is practically independent of the primary current frequency.

Our estimation of the loss power in a single sheet of 0.3 mm in thickness was performed on the basis of the expressions given in [16] and gave the value of about $10^{-3}$ W for the maximum $B_r$ = 0.035 T (corresponding to 1 A current in the primary coil) at the frequency of 120 Hz. The rod consists of 50 isolated sheets but the centrally located sheets have smaller losses: the radial component decreases towards the center. Averaging and summarizing over all the sheets, we estimate the maximum total loss power in the rod as 0.05 W at the current of 1 $A_{rms}$ . Taking into account that the losses are proportional to $B_r^2$, we evaluate the loss power in the core of the device with the ring in the superconducting state ($B_r$ = 0.003 T and the frequency of 120 Hz) as $5{\cdot}10^{-4}$ W. The losses are changed as the square of the frequency. As we shall see below, this value is negligible in the comparison with the AC losses in the superconductor. For example, at the primary current of 4 $A_{rms}$ and frequency of 120 Hz, the losses in the superconductor is 0.17 J per the period (see Fig. 4), or 20.4 W while the core losses at the same conditions are $8{\cdot}10^{-3}$ W.

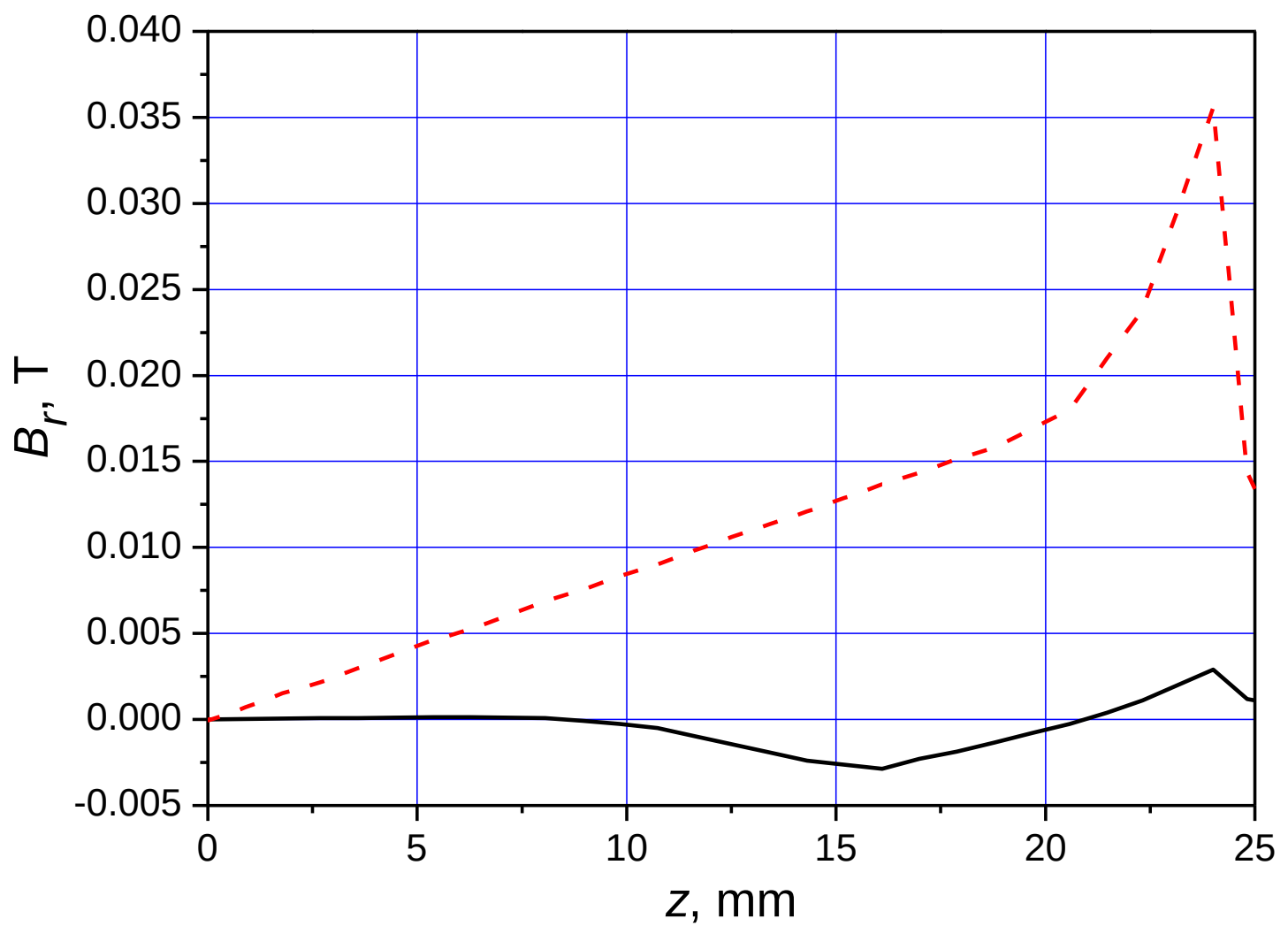

Fig. 2. Calculated radial component of the magnetic field density at the lateral surface of the ferromagnetic rod without (red dotted line) ring and with a well-conducting ring as a secondary (black solid line) at the primary current of 1 A. Due to symmetry it is shown a half of the rod.

**4. Measurement results**

The typical current oscilloscope traces of the primary current and magnetic field density are presented in Fig. 3a. The AC loss measurements were performed till quenching (transition of the ring into the non-superconducting state) characterized by appearance of distortions in the voltage waveform. The typical $B_\Sigma - i_1$ loop is shown in Fig. 3b. At a low current (<0.2 $A_{rms}$) the loop degenerates into a straight line thus showing that AC losses in the superconductor and magnetic core are less than the precision of the measurements.

The current and frequency dependences of AC losses per the period determined using the obtained waveforms of $i_1$ and $B_\Sigma$ , and Eq. (5) are shown in Figs. 4 - 6 for ring 2. The losses in ring 1 are about 10% lower and have the same dependences on the current and frequency.

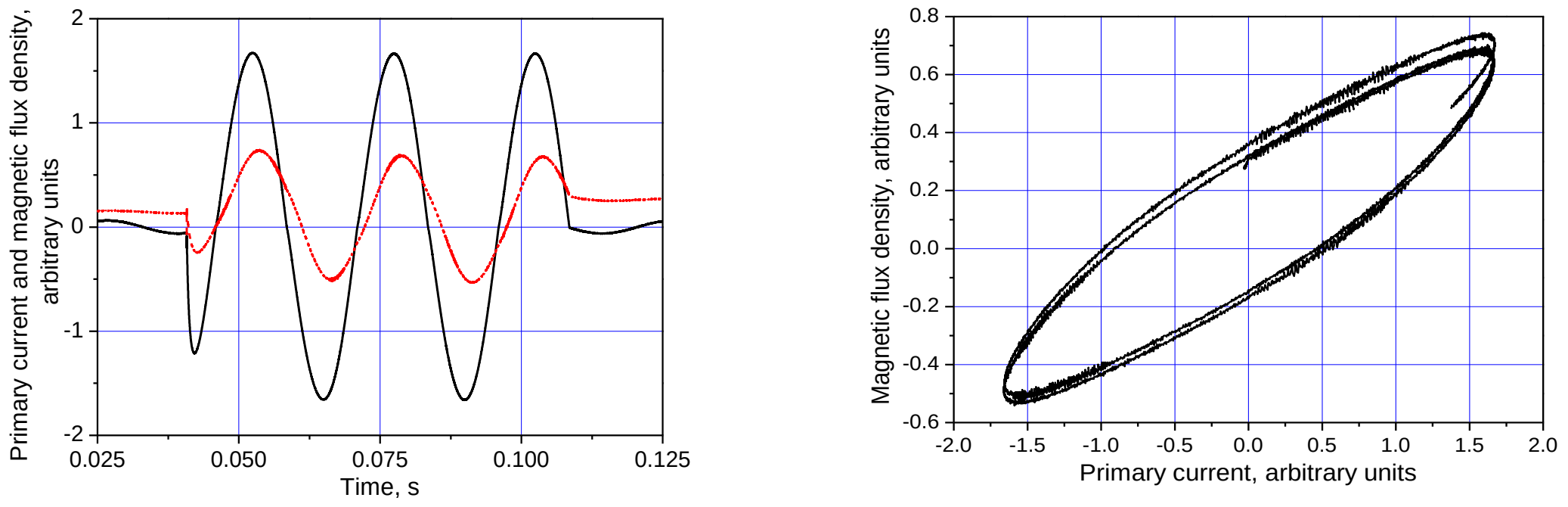

(a) (b)

Fig. 3. (a) Typical waveforms of the current (black solid curve and magnetic flux density (red dotted curve); (b) loop $B_\Sigma - i_1$. The frequency of the current is 40 Hz.

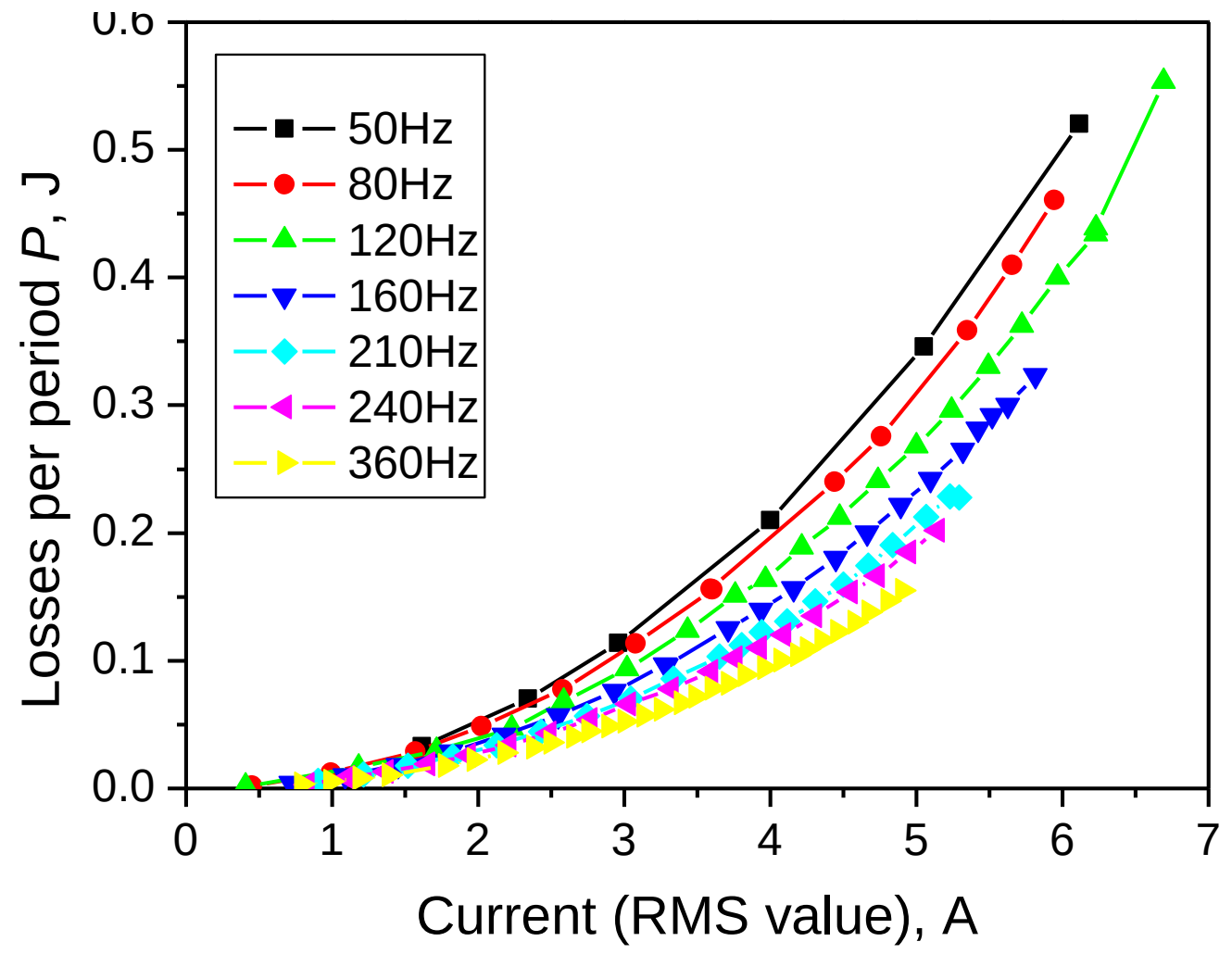

Fig. 4 Dependence of AC losses in the $MgB_2$ ring on the primary current at different frequencies and temperature of 4.2 K. Ring 2.

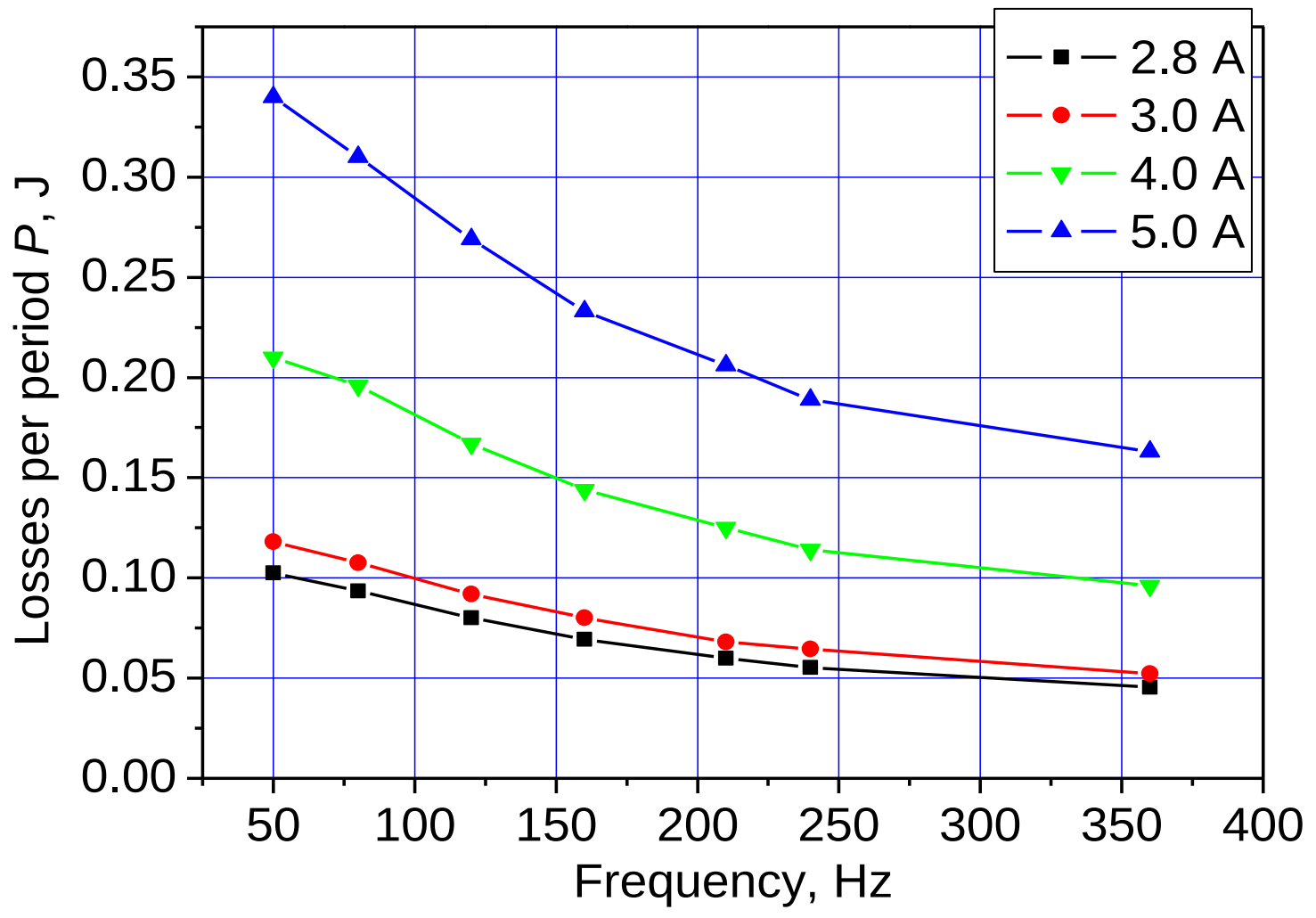

Fig. 5 Losses per period as a function of frequency at different rms currents in the primary coil and at temperature of 4.2 K. Ring 2.

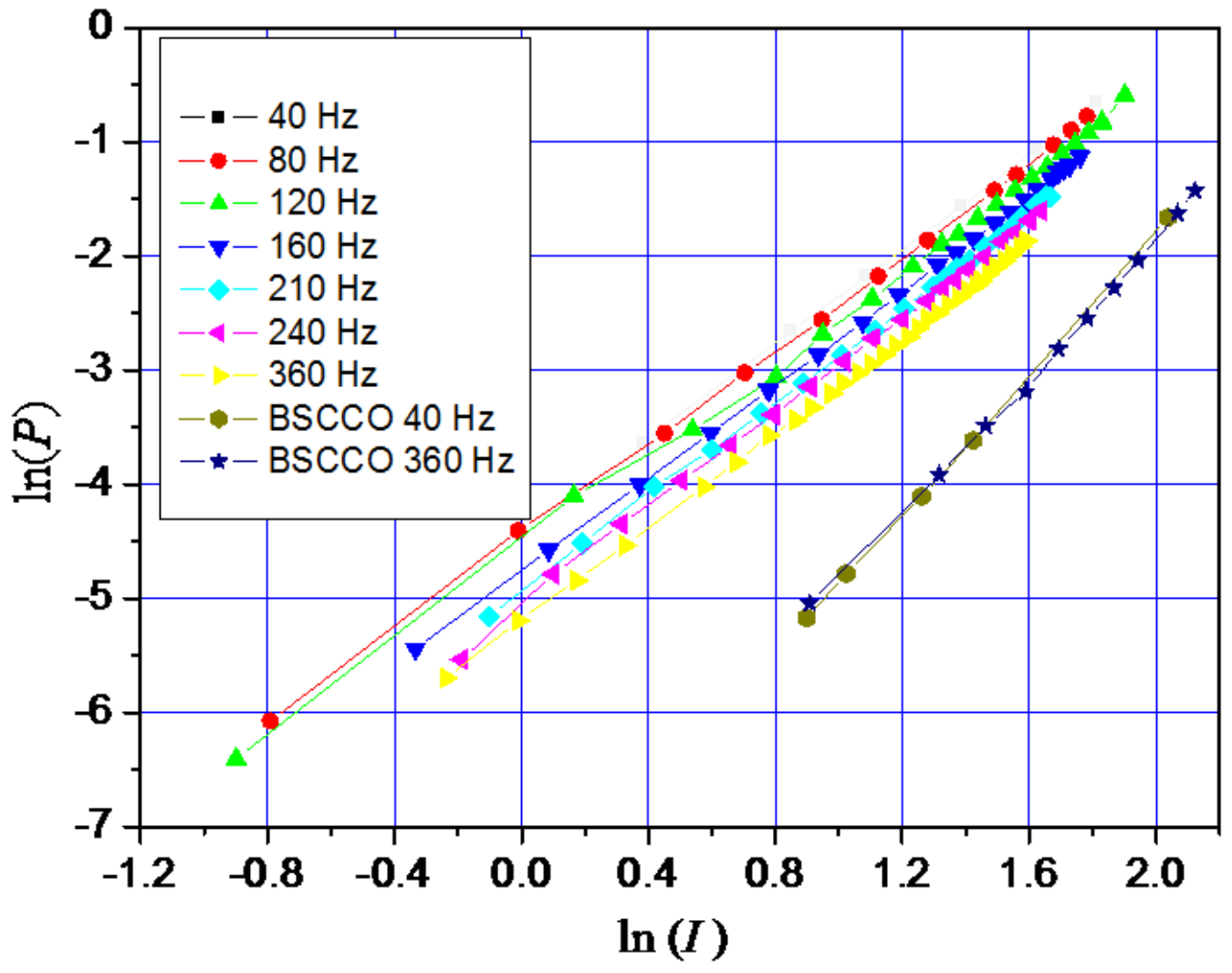

Fig. 6. Logarithm of AC losses in the $MgB_2$ ring and BSCCO cylinder as functions of logarithm of the primary current.

## 5. Discussion

The measurement results show that AC losses in the $MgB_2$ rings have the following features:

- loss value per the period depends strongly on the frequency ($\sim 1/f^{0.4}$ at 5 $A_{rms}$),
- dependences of losses on the primary current are well fitted by a power law $P \sim i^m$ with exponent $m$ of about 2.1 (Fig. 6).

The Bean model predicts AC losses independent of frequency and fitted by a power law with exponent of about 3 [11, 17]. The widely used model based on the power law E-J characteristic, $E \sim J^n$, gives deviations from the Bean model results, but, even at relative low index $n = 7 \div 9$, these deviations are not such pronounced. For comparison, we measured in a BSCCO cylinder using the same experimental technique and Hall sensor (experiment details are given in Appendix).

The dependence of AC losses in BSCCO on the primary current is fitted by a power law with exponent $m$ of about 2.8 (Fig. 6, curves for BSCCO). These results are in full accordance with theoretical and experimental investigations presented in [11, 12,18-20].

In order to obtain an adequate description of the E-J characteristic of the investigated samples, we have compared AC losses based on different models with Bean's model giving an exact analytical solution [11], calculating AC losses using COMSOL within the axial-symmetric 1D approximation where the primary coil, superconductor and core were infinitely long. Using the power law E-J characteristics with large indexes $n$ ($n > 15$), we obtain good accordance of the results with Bean's model. The simulation of the BSCCO cylinder with $n = 7$ gives $m$ of about 2.8. An attempt of using the power law E-J characteristic with lower indexes $n$ ($n < 5$) to simulate the AC loss behavior in the $MgB_2$ ring leads to $m \approx 2.5$ that is sufficiently higher than the value found experimentally for the $MgB_2$ rings.

To clarify the reasons of such behavior of AC losses in $MgB_2$, we measured the E-J characteristic of the samples applying pulses of a direct current to the primary coil. In our experimental set-up (Fig. 1), the AC source was replaced by a DC one. At low primary currents, when currents are induced in a ring is less than the critical value, the measured magnetic field $B_\Sigma$ is proportional to the primary current. The waveforms of the primary current and magnetic field are given in Fig. 7a for the case when the maximum current in the ring exceeds slightly the critical current value. The magnetic field increases linearly with the primary current until the ring current achieves the critical value (Fig. 7b). After this point the dependence becomes strongly non-linear: the magnetic field penetrates into the ring.

The current in the ring and voltage drop across it were determined using Eqs. (3) and (4). The voltage-current (V-I) characteristic of ring 2 obtained in this experiment is shown in Fig. 8 and demonstrates almost linear increase after the critical value. For the comparison, the calculated power-law V-I dependence with the index $n = 100$ and the same critical current determined by the criterion 1 μV/cm is presented on the plot (blue dotted line). Assuming that the superconductor is homogeneous and the current density is uniform over the superconductor cross-section, we obtain that the V-I characteristic is well described in the framework of the extended critical state model [21]:

$$E = \begin{cases} \rho_f [J - \mathrm{sign}(J) J_c], & |J| > J_c \\ 0, & |J| < J_c \end{cases} \qquad (9)$$

where $\rho_f$ estimated from the curve of Fig. 8 equals about $1.8\cdot10^{-9}$ Ω·m, the critical current density $J_c \cong 22000$ A/cm$^2$ and the critical current $I_c \cong 5500$ A for ring 2. For ring 1 these values are: $\rho_f = 10^{-9}$ Ω·m; $J_c \cong 21000$ A/cm$^2$ , and $I_c \cong 4500$ A.

Expression (9) well fits the E-J characteristics of low-temperature and some high-temperature superconductors [21-24] in which the creep flux can be neglected. The E-J characteristic is determined by the flux flow regime and, in the general case, the flux flow resistivity $\rho_f$ is a function of the local magnetic field and temperature [22, 23]. Note, that in the Bean model $\rho_f$ is assumed to be infinitely large.

The calculation results for the $MgB_2$ with the E-J characteristic in the form of (9) are shown in Fig. 9. For the considered approximation of infinitely long primary coil, superconducting cylinder and core, the magnetic field at the superconductor surface equals to the primary current multiplied by the turn number per length unit.

In contrast to Bean's model, the extended model based on (9) is characterized by an increase of the current density above the critical value. The relative increase of the current density $\Delta J/J_c$ due to a finite flux flow resistivity can be evaluated as

$$\frac{\Delta J}{J_c} \approx \frac{\omega \mu_0 H_0 \Delta}{\rho_f J_c} \le \frac{\omega \mu_0 H_0 d_{sc}}{\rho_f J_c}, \qquad (10)$$

where $d_{sc}$ is the characteristic size of a superconductor: its radius or thickness.

At the evaluation of AC losses in low temperature superconductors, one can neglect $\Delta J$ in wide range of the frequencies and applied magnetic field (the diameter of a superconducting filament is less than 10 μm or even 1 μm; the critical current density is order $10^{10}$ A/m$^2$). Therefore, low temperature superconductors are well described by Bean's model.

The thickness of the investigated $MgB_2$ samples is three orders higher. Therefore, when $\omega H_0$ is relatively low, the penetration depth $\Delta \approx H_0/J_c$ and the current density is close to the critical value. In this case the losses can be evaluated using Bean's model: the losses per the period are proportionally to $H_0^3$. With the increase of the applied magnetic field or/and frequency, the current density increases and can exceed the critical value several times. In this case the loss density is $p = EJ = \rho_f (J - J_c) J \approx \rho_f J^2$ and the properties of a superconductor are close to the properties of a normal metal where losses per the period are proportionally to $H_0^2$. In a normal metal with the thickness larger than the skin depth, the losses per the period are proportional to $1/\sqrt{f}$.

The calculation results given in Fig. 9 well agree with the analysis above: the fitting exponent for AC losses decreases from about 2.9 at low field till 2.2 at high field and frequency dependence is proportionally $1/f^{\ 0.38}$. These results are also in full accordance with our experiment. Note, that our results qualitatively agree with the field dependence of AC losses in a bulk MgB2 sample reported in [24]. Authors of the cited paper observed a gradual decrease of the exponent $m$ from 3 to 1 with the increase of magnetic field. The dependence of the AC losses on the field contains a long segment with $m$ = 2.

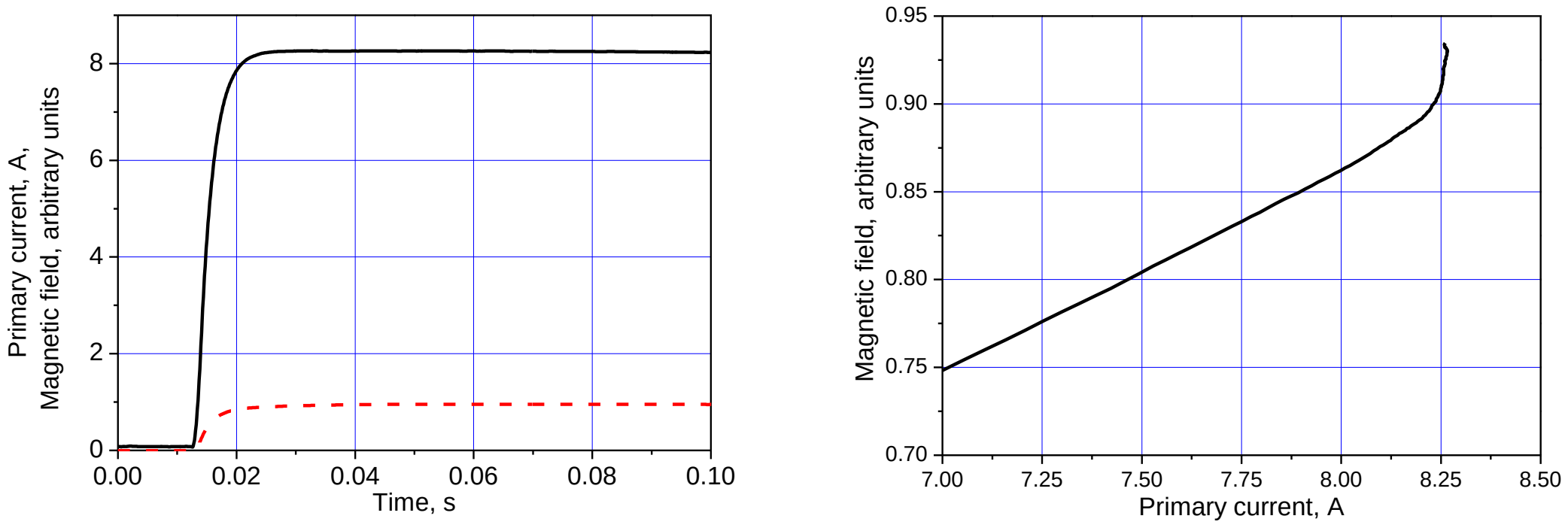

(a) (b)

Fig. 7 Current in the primary coil (black solid line) and magnetic field (red dashed line) in the DC pulse experiment (a); magnetic field vs. primary current (b).

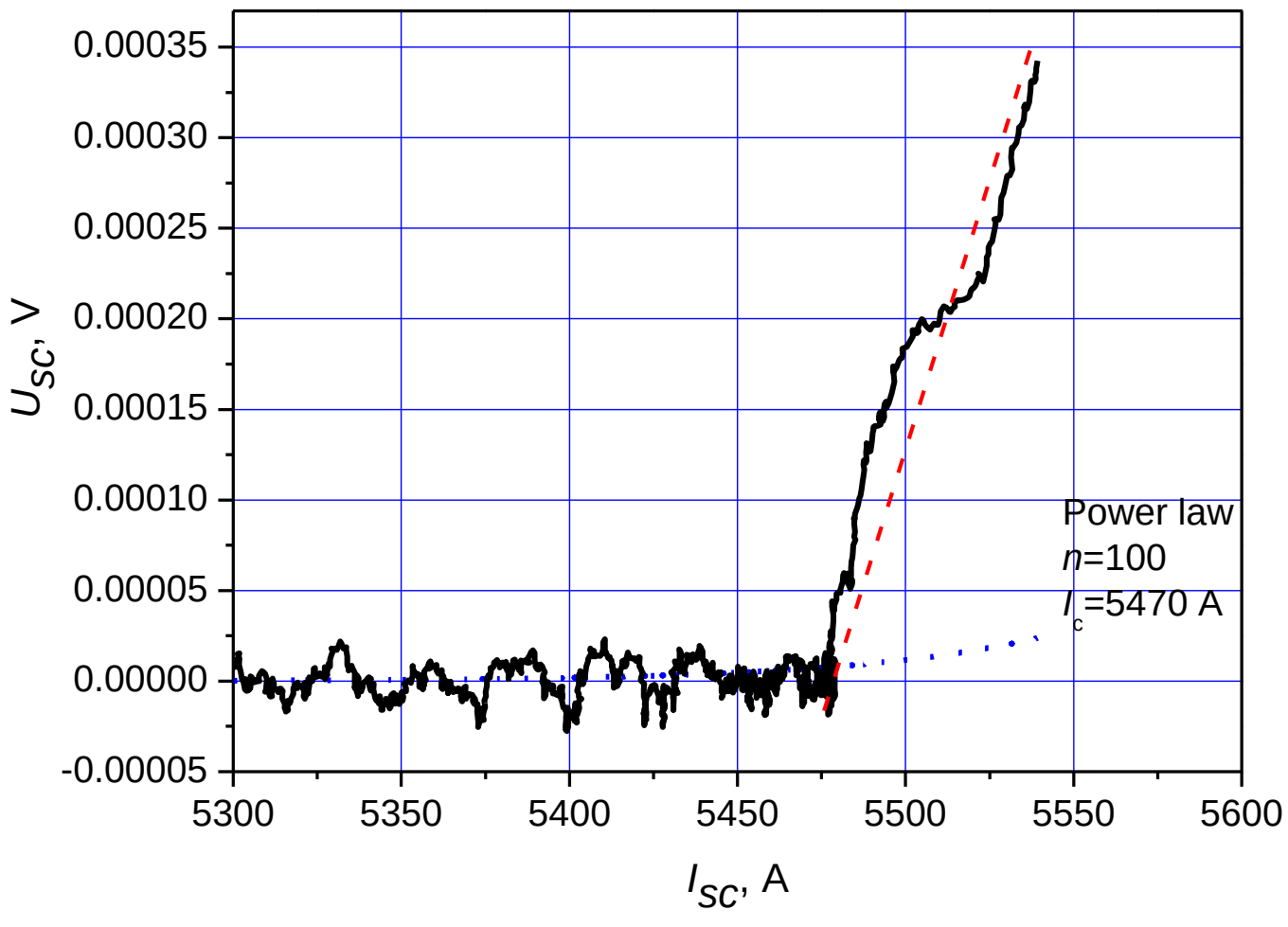

Fig. 8 V-I characteristic of $MgB_2$ ring.

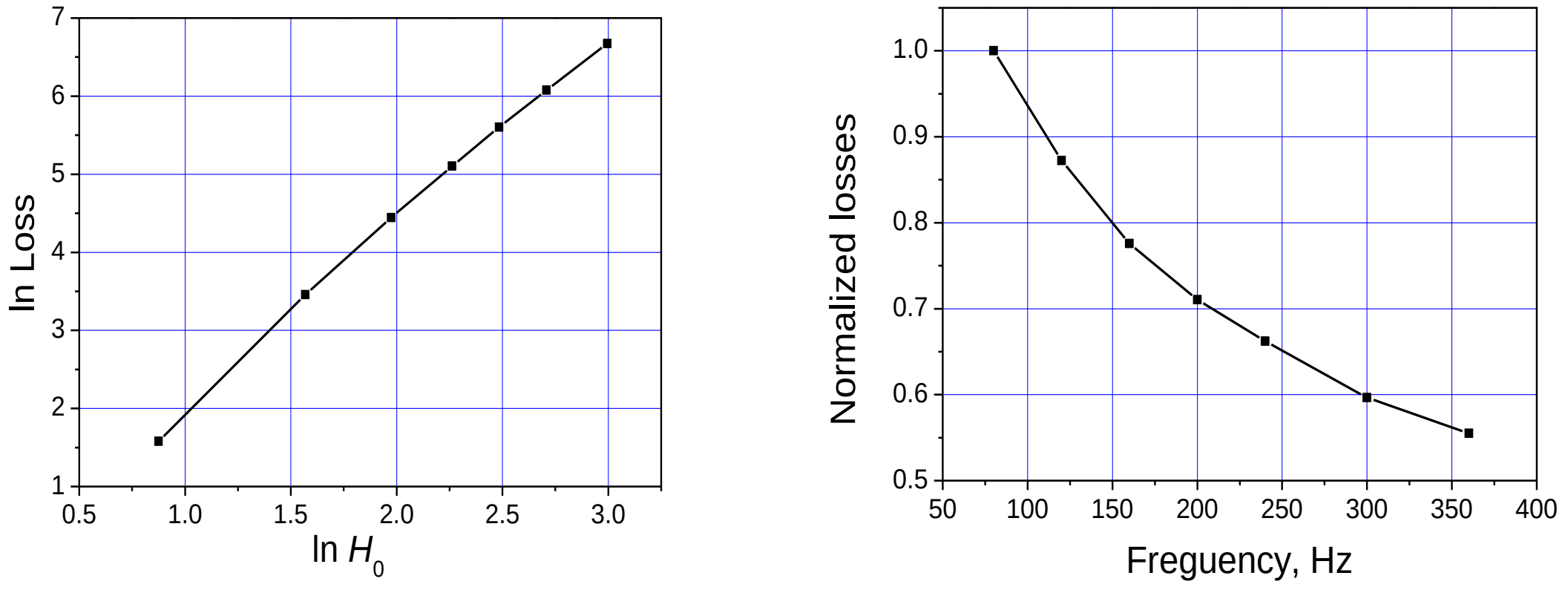

a) b)

Fig. 9 Calculation using the extended critical state model in the axial-symmetric 1D approximation at $\rho_f$ =$10^{-9}$ Ω·m and $f$ = 100 Hz: a) logarithm-logarithm loss dependence; b) the frequency dependence at $H_0$=1.2·$10^5$ A/m, losses are normalized by their value at $f$ = 80 Hz.

## 6. Conclusion

The application of the transformer technique allowed us to measure AC losses and E-J characteristic of the bulk $MgB_2$ samples with an induced transport current. The dependence of AC losses on the current is well fitted by the power law with the exponent of about 2.1. The losses per the period of the current decrease with the increase of the frequency $f$ as $1/f^{0.4}$. The measured E-J characteristic is well described by the extended critical state model, which gives the loss dependences on the magnetic field and frequency close to the experimental results.

## Appendix

The BSCCO cylinders were fabricated using the melt cast technology developed by Hoechst [25,26]. The melt cast process relies on casting a homogeneous melt of the starting materials in moulds of the required shape and size. The hot melt is cast into rotating moulds, where it is evenly distributed on the inner side of the walls. After solidification of the melt, elements undergo a suitable heat treatment in order to obtain proper superconducting properties. The obtained casts were machined into 30 mm height, 35 mm inner diameter and 5.1 mm wall thickness cylinders. Characterization of the specimens by DC four-point method gave the value of 570 A/cm$^2$ for the critical current density from 1 μV/cm criterion and the critical temperature of 94 K. DC current– voltage characteristic and critical current vs. magnetic field have been studied in detail and described elsewhere [26]. Using the transformer experimental configuration we have studied the response of the cylinders to a step of the current in the primary coil [12]. For current densities above about 600 A/cm$^2$ the E-J characteristic is well fitted by the power law with the power index equal to 7.

The BSCCO cylinder forms a single-turn secondary coil. It was placed inside a cooper 400 turn primary coil of a two-coil transformer. The primary coil with the internal diameter of 38 mm and the height of 38 mm was wound from 0.3 mm diameter cooper wire. A ferromagnetic laminated rod assembled from the transformer steel of 0.3 mm thickness with an effective diameter of 20 mm and height of 48 mm was inserted into the cylinder.